\documentclass{aa}  

\usepackage{graphicx}
\usepackage{txfonts}

\usepackage{graphics}
\usepackage{float}
\usepackage{hyperref}
\usepackage{natbib}
\usepackage{gensymb}

\begin{document}

  \title{A catalog of stereo anaglyphs of the nucleus of comet 67P/Churyumov-Gerasimenko}

  \author{Philippe Lamy \inst{1} \and Guillaume Faury \inst{2} \and David Romeuf \inst{3} \and Olivier Groussin \inst{4} \and Joelle Durand \inst{5} \and Laurent Beigbeder \inst{6} \and Kea Khum \inst{6}}

  \institute{Laboratoire Atmosph\`eres, Milieux et Observations Spatiales, CNRS \& Universit\'e de Versailles Saint-Quentin-en-Yvelines,
11 boulevard d'Alembert, 78280 Guyancourt, France 
  \and AKKA Informatique et Syst\`emes, 6 rue Roger Camboulives, BP 13633, 31036 Toulouse cedex 1, France 
  \and Universit\'e Claude Bernard Lyon 1, 43 bd du 11 novembre 1918, 69622 Villeurbanne, France 
  \and Aix Marseille Universit\'e, CNRS, CNES, LAM, Marseille, France
  \and CNES-Science Operations and Navigation Center, avenue Edouard Belin, 31506 Toulouse, France 
  \and GFI Informatique, 1 Rond-point du G\'en\'eral Eisenhower, 31100 Toulouse, France}

  \abstract
   {The OSIRIS/NAC camera aboard the Rosetta spacecraft has acquired approximately 27000 images of comet 67P/Churyumov–Gerasimenko at spatial scales down to a few centimeters.}
   {We exploit the numerous sequences of images separated by a few minutes suitable for stereo reconstructions to produce anaglyphs offering three-dimensional views complementary to other technics and particularly suitable to understand the detailed topography of the nucleus.}
   {Starting from the calibrated images, a selection is performed of suitable pairs on the basis of their angular separation (parallax).
	Anaglyphs are then constructed using standard softwares.
	Each anaglyph is documented by a set of 17 parameters which provide the contextual information. 
	}
   {The overall collection includes 1823 anaglyphs at all scales from the coma down to a few centimeters.   
	They are available on a dedicated website at the following URL: https://rosetta-3dcomet.cnes.fr and can be searched using the associated parameters.}
   {}
	

  \maketitle

\section{Introduction}

Three-dimensional rendering of the shape and surface of minor bodies of the solar system is essential to the understanding of their formation, to the characterization of their surface topography and to the study of their surface processes.
In the context of the early space missions, emphasis was given to the shape reconstruction by stereo techniques to obtain quantitative information and topographic measurements.  
The planetary group at Cornell University was at the forefront of this effort with the development of a method that combines several hundred stereo control points, limbs and terminators constraints, and light curves information \citep{Simonelli1993}.
It has been applied to several asteroids and cometary nuclei, see numerous publications by P.C. Thomas and coworkers, for example \citet{Thomas1999} for asteroid Mathilde and \citet{Thomas2013} for the nucleus of comet 103P/Hartley~2.
Later developments include the stereophotoclinometry method by \citet{Gaskell2008} and the multi-resolution photoclinometry by deformation method by \citet{Capanna2013}
In particular, the latter development was specifically undertaken in the framework of the Rosetta mission to comet 67P/Churyumov-Gerasimenko.

Anaglyphs have seen a recent resurgence probably due to the need of conveniently presenting three-dimensional (3D) images to a general audience on computer screens via the Web, CDs, and DVDs.
As a consequence, space agencies (NASA in particular) and space imager teams have released hundred's of anaglyphs of solar system objects conveniently archived on dedicated websites.
Planet Mars is an excellent example with anaglyphs from Mars Pathfinder\footnote{\url{https://mars.nasa.gov/MPF/mpf/anaglyph-arc.html}} and from the Curiosity and Opportunity rovers\footnote{\url{https://mars.nasa.gov/mars3d/}}.
Other examples are available encompassing asteroids (e.g., Vesta), and planetary satellites (e.g., the Moon, Enceladus and its geysers.), often seen as ``APOD'' (Astronomical Picture of the Day)\footnote{\url{https://apod.nasa.gov/apod/}}.

It has recently been realized that anaglyphs have their own merits for scientific analysis.
For instance, the HiRISE team\footnote{\url{https://hirise.lpl.arizona.edu/}} has placed a major emphasis on stereo imaging needed to make small-scale
topographic measurements, essential both to the characterization of candidate landing sites and to the quantitative study of surface processes. 
During the first two years of operations ending November 2008, nearly 1000 stereo pairs have been acquired to produce anaglyphs and they are presented on equal terms with digital terrain models (DTMs) on their website thus emphasizing their complementary nature.
DTMs do present the advantage of offering quantitative measurements of topography but they require many favorable view angles and a time-intensive process. 
They involve sophisticated software and a lot of time, both computing time and human operator time.  
On the contrary, anaglyphs require only two matching views and are fairly straightforward to generate.
They also preserve the small-scale texture and somehow lead to an increase of the spatial resolution when the brain recombines the two images.

The OSIRIS/NAC camera \citep{Keller2007} aboard the Rosetta spacecraft has acquired approximately 27000 images at different distances and therefore at different spatial scales of comet 67P/Churyumov–Gerasimenko, from global images of the bi-lobed nucleus and its jets down to topographic details of a few centimeters on different regions of the nucleus. 
Although the potential of stereo anaglyphs was not realized during the preparation of the observational program, many images have been obtained in sequences with time intervals of a few minutes so that the rotation of the nucleus and the displacement of the spacecraft allow the a-posteriori selection of suitable pairs to construct anaglyphs. 
In fact, several published articles have already included anaglyphs to support their scientific analysis, for instance \citet{Auger2015}, \citet{Mottola2015}, and \citet{Pajola2017}.
We realized that a systematic production of anaglyphs and their documented presentation on a dedicated website would constitute a relevant tool to future detailed analysis of the nucleus of 67P, would renew the interest of the general public, and would provide a valuable contribution to the legacy of the Rosetta mission.
This present article reports on this effort and its outcome.
After briefly summarizing the Rosetta mission (Section~1), the OSIRIS/NAC instrument (Section~2), and the processing of the images (Section~3), we describe the construction of the anaglyphs and display a few examples (Section~4).
Section~5 describes their parameters and (Section~6) the on-line catalog.
We conclude in Section~7.

\section{The Rosetta mission}
\begin{figure}[htpb!]
   \centering
   \includegraphics[width=0.5\textwidth]{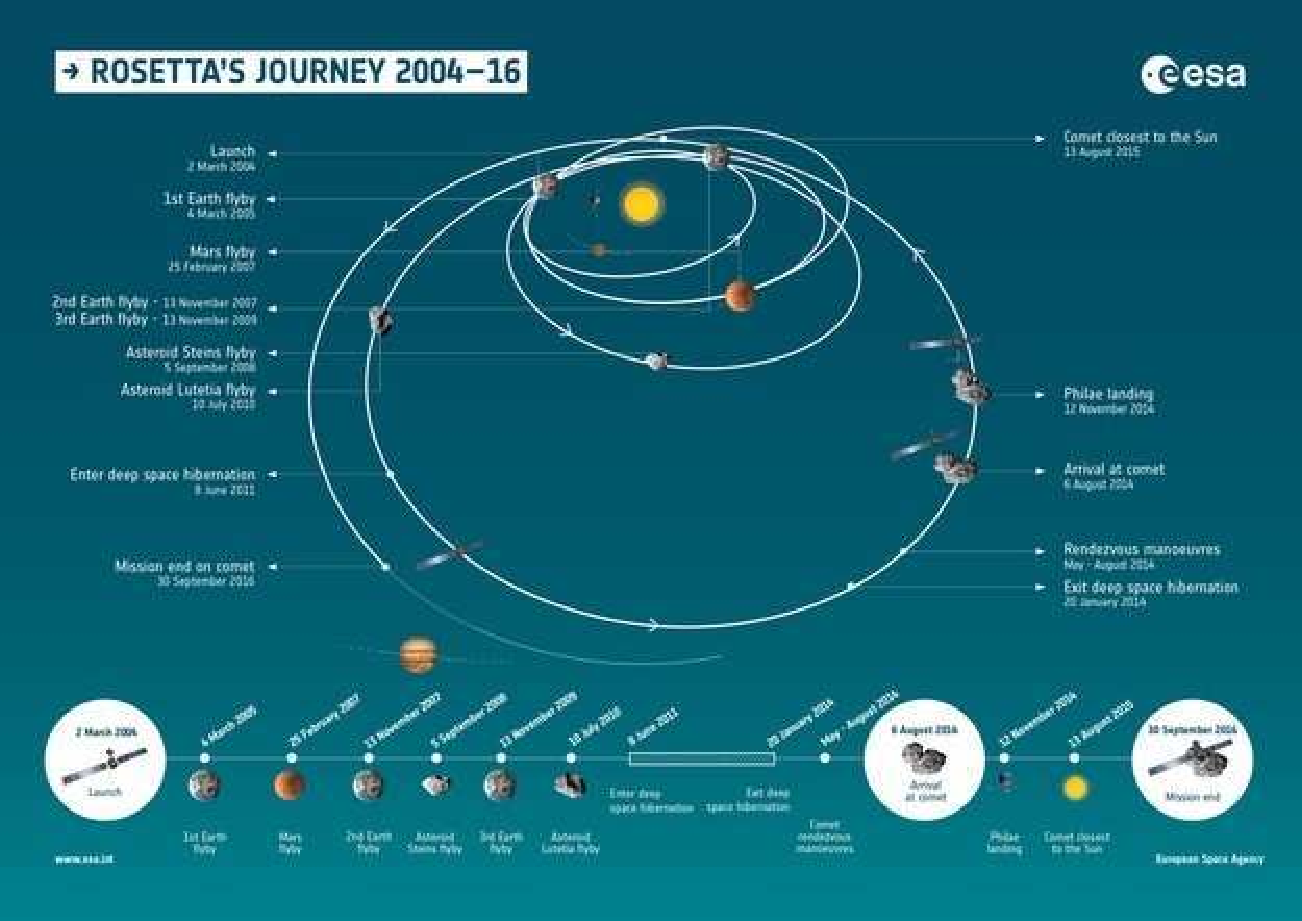}
   \caption{Interplanetary trajectory of the ROSETTA spacecraft (courtesy ESA).}
   \label{traj}
   \end{figure}

The International ROSETTA Mission was the planetary cornerstone mission in ESA's long-term program “Horizon 2000” approved in November 1993.
It was a cooperative project between ESA, several European national space agencies, and NASA.
Its main scientific objective was the investigation of the origin of our solar system by performing in-situ observations of a comet, a member of a family of objects thought to be the most primitive. 
The mission was conceived as a rendezvous with its target comet while inactive at a large heliocentric distance so as to allow studying its nucleus, followed by an escort phase to and past perihelion to characterize the development of cometary activity. 
Trajectory analysis indicated that it was possible to fly by up to two asteroids during the journey to the comet. 
Final selection, once the launch date was firmly established, led to choosing comet 67P/Churyumov–Gerasimenko as the main target of the rendezvous and asteroids (2867) Steins and (21) Lutetia as flyby targets (Fig.~\ref{traj}). 
The spacecraft consisted of two mission elements, the ROSETTA orbiter and the ROSETTA lander PHILAE altogether comprising a suite of 21 scientific instruments (\citet{Glassmeier2007}).

After its launch on 2 March 2004 with an Ariane V rocket, the spacecraft started its long journey to the comet.
Four gravity assists, three by Earth and one by Mars were required to propel the ROSETTA spacecraft to a distance of 3.6 AU to rendezvous with the comet. 
After the second flyby over asteroid (21) Lutetia on 10 July 2010, the spacecraft was put in a state of deep space hibernation during 31 months and woken up on 20 January 2014. 
It arrived at a distance of 100 km from the nucleus on 6 August 2014 and then started a complex journey around the nucleus in order to fulfill its scientific mission.

\begin{figure}[htpb!]
   \centering
    \includegraphics[width=0.5\textwidth]{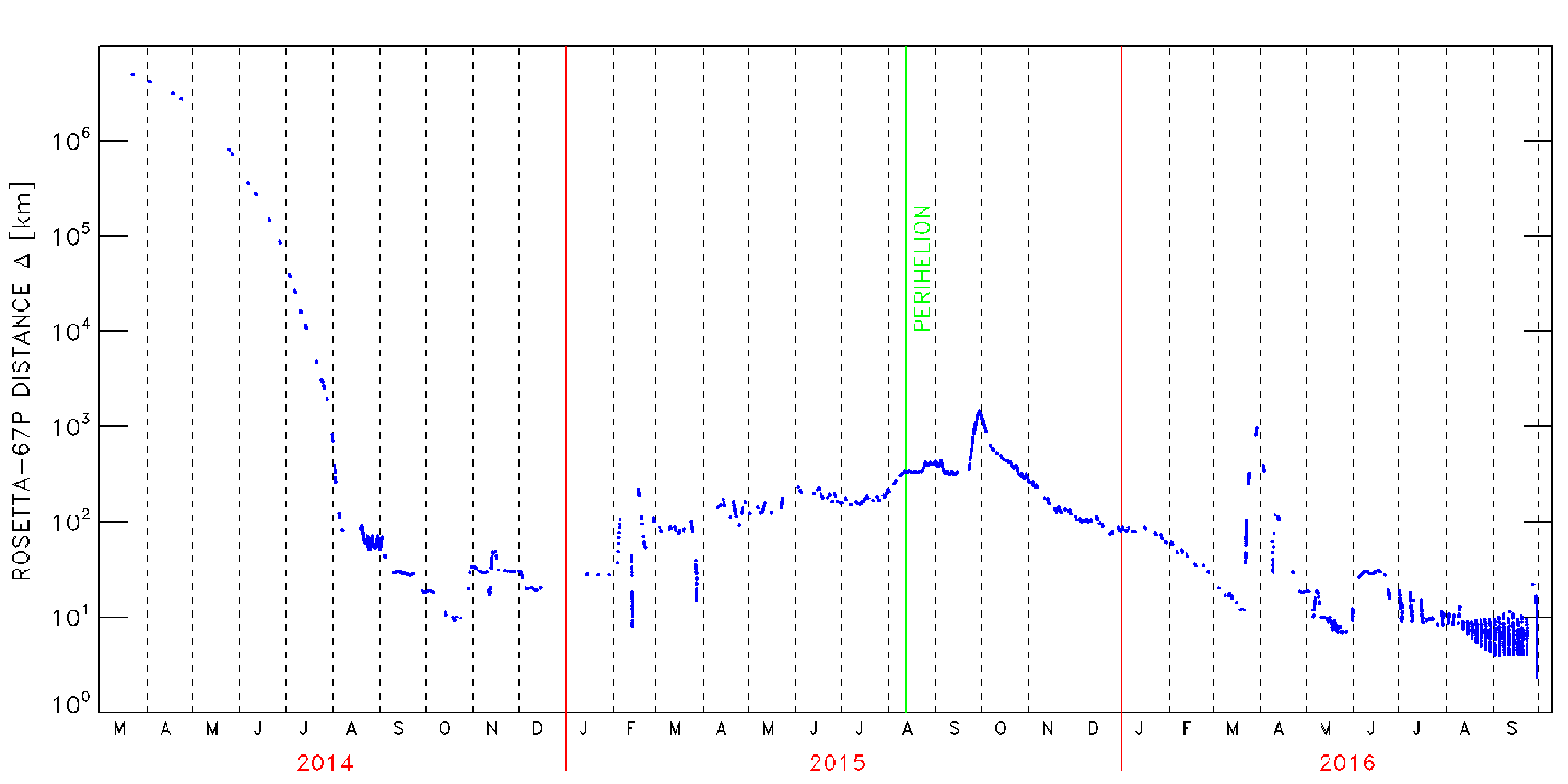}
   \caption{Evolution of the distance (km) between the ROSETTA spacecraft and the nucleus of comet 67P/Churyumov–Gerasimenko from March 2014 to September 2016 using a logarithmic scale. 
	The blue dots corresponds to images taken by the OSIRIS/NAC camera for a total amount of 26827 images.}
    \label{dist_rosetta_chury}
\end{figure}

Various constraints besides scientific dictated this circum-cometary navigation, particularly the safety of the spacecraft. 
Consequently, its distance to the nucleus varied considerably as illustrated in Fig. ~\ref{dist_rosetta_chury} and conspicuously increased around perihelion time as a measure of protection against the increasing cometary activity. 
 
\section{The OSIRIS-NAC instrument}

\begin{figure}[htpb!]
   \centering
   \includegraphics[width=0.5\textwidth]{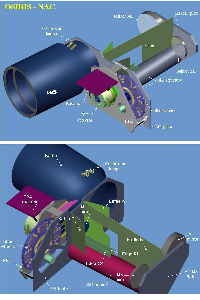}
   \caption{Two schematic views of the OSIRIS Narrow Angle Camera.}
              \label{osiris_shem}%
\end{figure}

\begin{figure}[htpb!]
   \centering
    \includegraphics[width=0.5\textwidth]{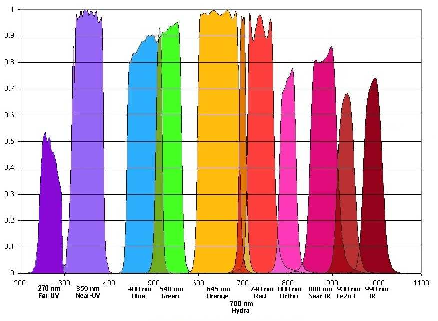}
   \caption{Spectral transmission curves of the 11 filters of the OSIRIS-NAC.}
   \label{osiris_transmission}%
\end{figure}

OSIRIS, the Optical, Spectroscopic, and Infrared Remote Imaging System of the ROSETTA mission (OSIRIS) consisted of a Narrow Angle Camera (NAC) and a Wide Angle Camera (WAC) operating in the visible, near infrared and near ultraviolet wavelength ranges (\citet{Keller2007}). 
OSIRIS was built by a consortium of 
the Max-Planck-Institut f\"ur Sonnensystemforschung, Gottingen, Germany, 
CISAS-University of Padova, Italy, 
the Laboratoire d’Astrophysique de Marseille, France, 
the Instituto de Astrofísica de Andalucía, CSIC, Granada, Spain, 
the Scientific Support Office of the European Space Agency, Noordwijk, Netherlands,
the Instituto Nacional de T\'ecnica Aeroespacial, Madrid, Spain, 
the Universidad Polit\'echnica de Madrid, Spain, 
the Department of Physics and Astronomy of Uppsala University, Sweden, 
and the Institut für Datentechnik und Kommunikationsnetze der Technischen Universität Braunschweig, Germany.
The NAC telescope was conceived and developed by the Laboratoire d’Astrophysique de Marseille in partnership with ASTRIUM (Toulouse) and with the European institutes of the consortium. 
Its optical concept implemented the three-mirror anastigmat (TMA) solution (Fig. ~\ref{osiris_shem} ) insuring low stray light by eliminating the central obscuration present in axial designs. 
The optimized solution required only two aspheric mirrors, the tertiary remaining spherical. 
It ensured a flat field, an axial pupil and an off-axis field of view of $2.20 \degree \times 2.22 \degree $. 
The telescope was equipped with a $2048 \times 2048$ pixel backside illuminated CCD detector with a UV optimized anti-reflection coating. 
The pixel size of 13.5~$\mu$m yielded an image scale of 3.9 arcsec/pixel. 
Two filter wheels holding 11 filters allowed the selection of spectral windows over the 250 to 1000 nm wavelength range (Fig.~\ref{osiris_transmission}).
Three of them were combined with a neutral density filter thus offering a total of 13 filters (Table ~\ref{filters_properties}). 
The images used to build the anaglyphs were obtained with these 13 filters labeled F16, F22, F23, F24, F27, F28, F41, F51, F61, F71, F82, F84, F88.

\begin{table}
\caption{Spectral properties of the NAC filters.}
\label{filters_properties}
\centering
\begin{tabular}{lccc}
\hline\hline
Code & Name & $\lambda_{c}$ & $\Delta \lambda$ \\
\hline
F16 & Near UV & 367.6 & 36.7\\
F22 & Orange & 648.6 & 85.2\\
F23 & Green & 536.4 & 64.5\\
F24 & Blue & 479.9 & 74.0\\
F27 & Hydra & 701.2 & 22.0\\
F28 & Red & 742.4 & 63.2\\
F41 & Near IR & 880.1 & 63.6\\
F51 & Ortho & 804.7 & 40.6\\
F61 & Fe2O3 & 931.7 & 34.7\\
F71 & IR & 986.1 & 34.5\\
F82 & F22 + Neutral density filter & & \\
F84 & F24 + Neutral density filter & & \\
F88 & F28 + Neutral density filter & & \\
\hline
\end{tabular}
\tablefoot{$\lambda_{c} =$ central wavelength, $\Delta \lambda = $ bandwith (nm) }
\end{table}

\begin{figure}[htpb!]
   \centering
   \includegraphics[width=0.5\textwidth]{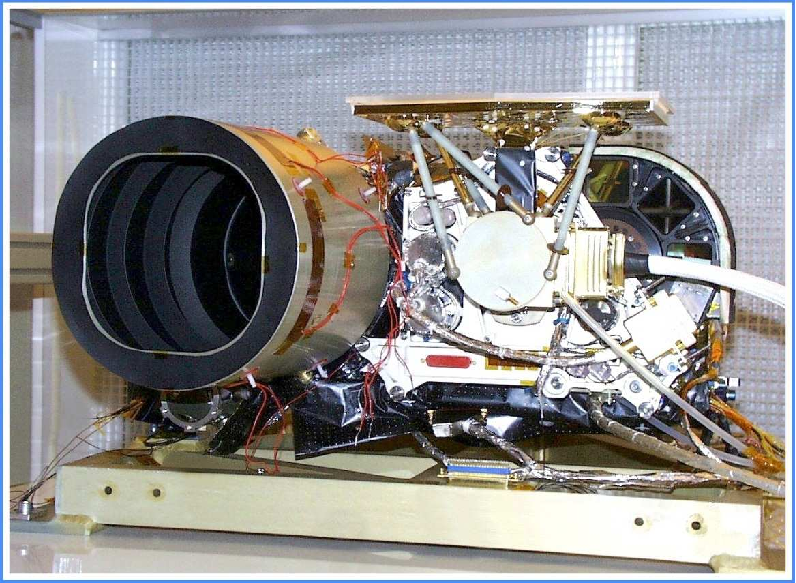}
   \caption{Flight model of OSIRIS-NAC.}
              \label{osiris-nac}%
\end{figure}

A view of the flight model of the NAC (without its thermal blanket) during final testing at Laboratoire d’Astropysique de Marseille is shown in Fig. ~\ref{osiris-nac}.

\section{The processing of the OSIRIS images}

\begin{figure}[htpb!]
   \centering
   \includegraphics[width=0.5\textwidth]{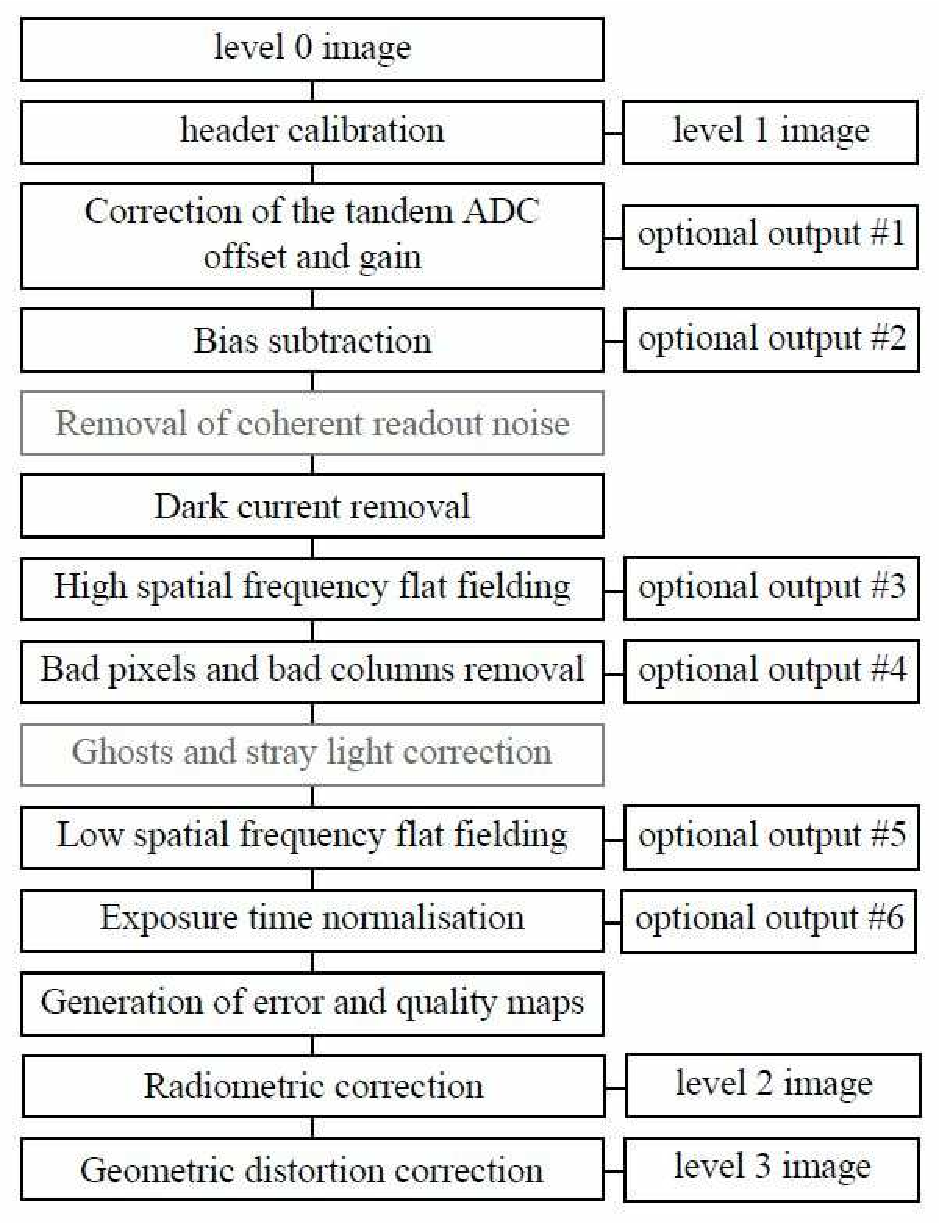}
   \caption{Flowchart of the OSIRIS calibration pipeline (reproduced from Fig. 2 of \citet{Tubiana2015}).}
              \label{osiris-pipe}%
\end{figure}

All OSIRIS images went through a complex, two-step calibration pipeline. 
The first step converted the raw data downloaded from the spacecraft to level-1 images with calibrated hardware parameters and S/C pointing information. 
The second step converted these level-1 images to radiometric calibrated and geometric distortion corrected images (level 2 and 3, respectively) through a series of successive tasks as illustrated in the flowchart of Fig. ~\ref{osiris-pipe} (\citet{Tubiana2015}). 
An additional level-4 was later implemented to associate to each pixel of each image its cometo-centric coordinates (longitude and latitude) so as to precisely localize the images on the nucleus.

\section{The construction of the stereo anaglyphs}

The potential of direct stereoscopic views of the nucleus of comet 67P/Churyumov–Gerasimenko was never realized during the preparation of the observational program of the NAC. 
The emphasis was put on the construction of numerical 3D models and digital terrain models (DTMs) since they provide quantitative information. 
Fortunately, it turned out that many observational programs required sequences of concatenated images so that a large number of image pairs were subsequently found suitable for anaglyph construction once their potential for scientific analysis was assessed. 
Optimal conditions for stereo imaging require two identical cameras spaced for stereo base and taking images at the same time to insure identical illumination conditions. 
Evidently, these optimal conditions were not met and with only one high resolution camera, the NAC, the stereo effect had to rely on its displacement. 
In practice, two effects come into play: the motion of the spacecraft and the intrinsic rotation of the nucleus, they relative importance depending upon the distance between the camera and the nucleus. 
The displacement is not arbitrary and must reproduce human vision with a stereoscopic base of 7 cm, the typical distance between our two eyes. 
This distance is adequate to introduce projection differences, so-called parallax, between 30 cm and several hundred meters. 
A horizontal parallax commonly considered as comfortable for the brain is $\approx$2$\degree$ corresponding to a base of $\approx 1/30$ of the distance to the near object, values which are adapted to the focal length of our eyes. 
After pre-selecting pairs of images on the basis of their time interval, the parallax between the two images was estimated and served as a criterion for the final selection. 
In practice, the criterion was relaxed to a range of parallax adapted to different conditions of visualization of the resulting anaglyphs. 
A parallax of $\approx$2$\degree$ is appropriate to viewing anaglyphs on a computer screen. 
A parallax of $\approx$0.5$\degree$, that is a base of 1/100, is suitable to a projection on a large screen at a conference. 
A parallax of $\approx$3.8$\degree$, that is a base of 1/15, is adapted to smartphones. 
A few anaglyphs with a parallax larger than 4$\degree$ (base $<$ 1/10), a limit beyond which the depth becomes too dilated, have been kept when presenting a specific interest. 
As mentioned in the next section, the value of the base is included in the name of the anaglyph and the parallax is given as a parameter in the database.

It has been pointed out above that part of the displacement results from the rotation of the nucleus. 
This has an adverse effect of modifying the cast shadows between the two images thus producing an uncomfortable incoherence perceived as a ``vibration'' between the limits of the two shadows. 
This effect can be removed by darkening regions so as to cover the maximum extent of the shadows recorded on both images. 
We have done so manually on a limited number of anaglyphs of particular value.
A systematic correction would be time-consuming and would in fact require the development of an automatic procedure, a task beyond the scope of the present catalog. 

Once a pair of suitable images has been selected, the construction of the anaglyph proceeds in several steps.

\begin{itemize}
\item A thresholding is applied to both images to adapt them to a common range of brightness (same minima and maxima). 
This operation is performed with the FITS Liberator software. 
\item The images are rotated using an image editor software so that their relative displacement is horizontal. 
\item The anaglyph is then created using the StereoPhoto Maker software\footnote{\url{http://stereo.jpn.org/eng/stphmkr/}} of Masuji Suto.
\item The anaglyph is finalized using Photoshop or GIMP to perform cosmetics correction, cutting out non-overlapping sections and applying sharpening filters as appropriate. 
Incoherent shadow limits may also be corrected as explained above.
\end{itemize}

\begin{figure*}[htpb!]
   \centering
   \includegraphics[width=0.9\textwidth]{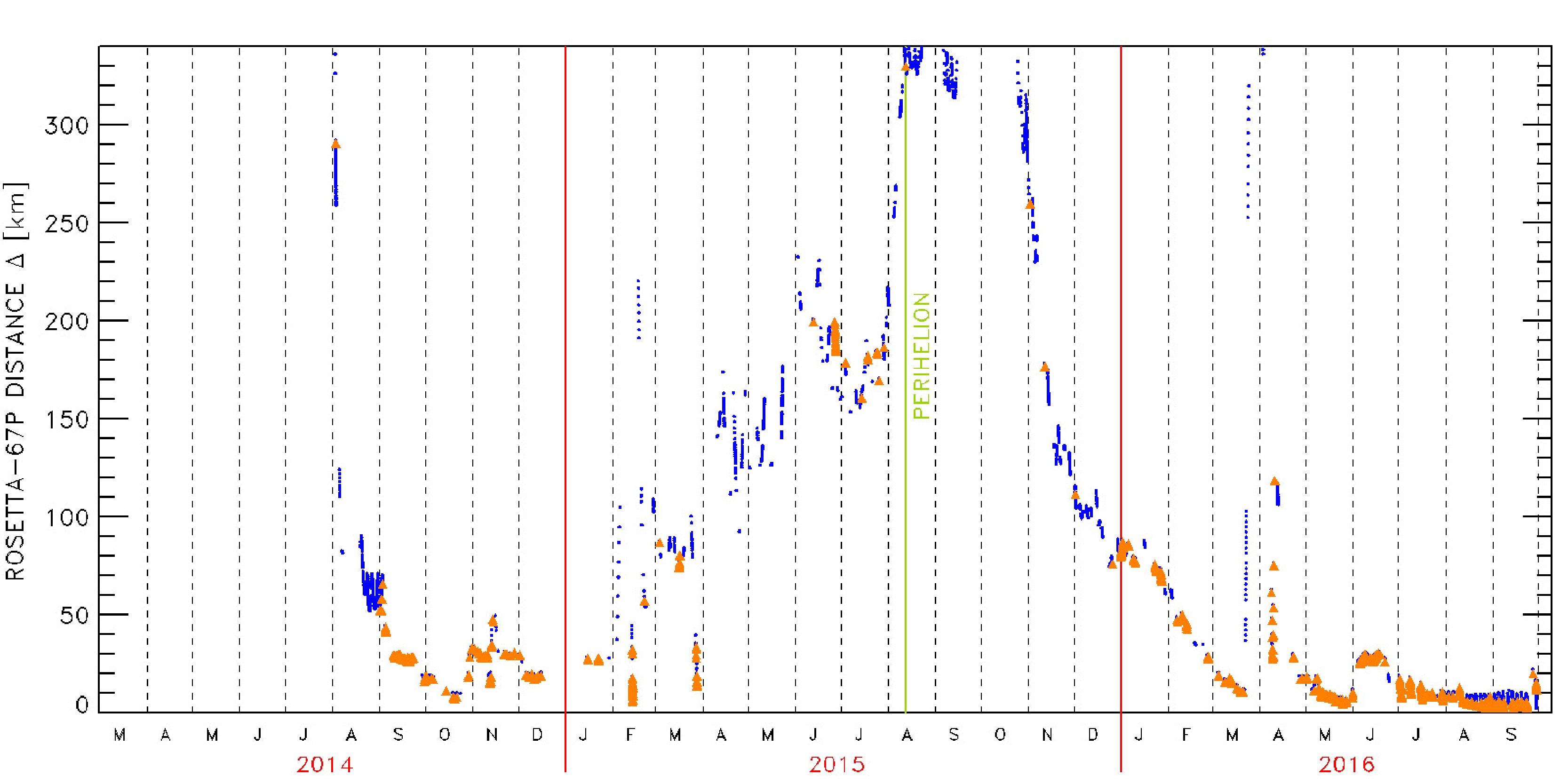}
   \caption{Same as Fig. ~\ref{dist_rosetta_chury} except that i) the distance scale is linear and limited to 350 km and ii) the constructed anaglyphs are superimposed as orange triangles.}
    \label{dist_rosetta_chury_closer}%
\end{figure*}

The very first anaglyph was constructed from images obtained on 3 August 2014. 
The subsequent temporal distribution is illustrated in Fig. ~\ref{dist_rosetta_chury_closer}. 
At time of writing, 1325 anaglyphs have been produced after analyzing a subset of 9921 NAC images. 
It is estimated that a total of 3000 to 4000 anaglyphs could be produced once the whole set of NAC images will be scrutinized.

\begin{figure*}[htpb!]
   \centering
   \includegraphics[width=0.9\textwidth]{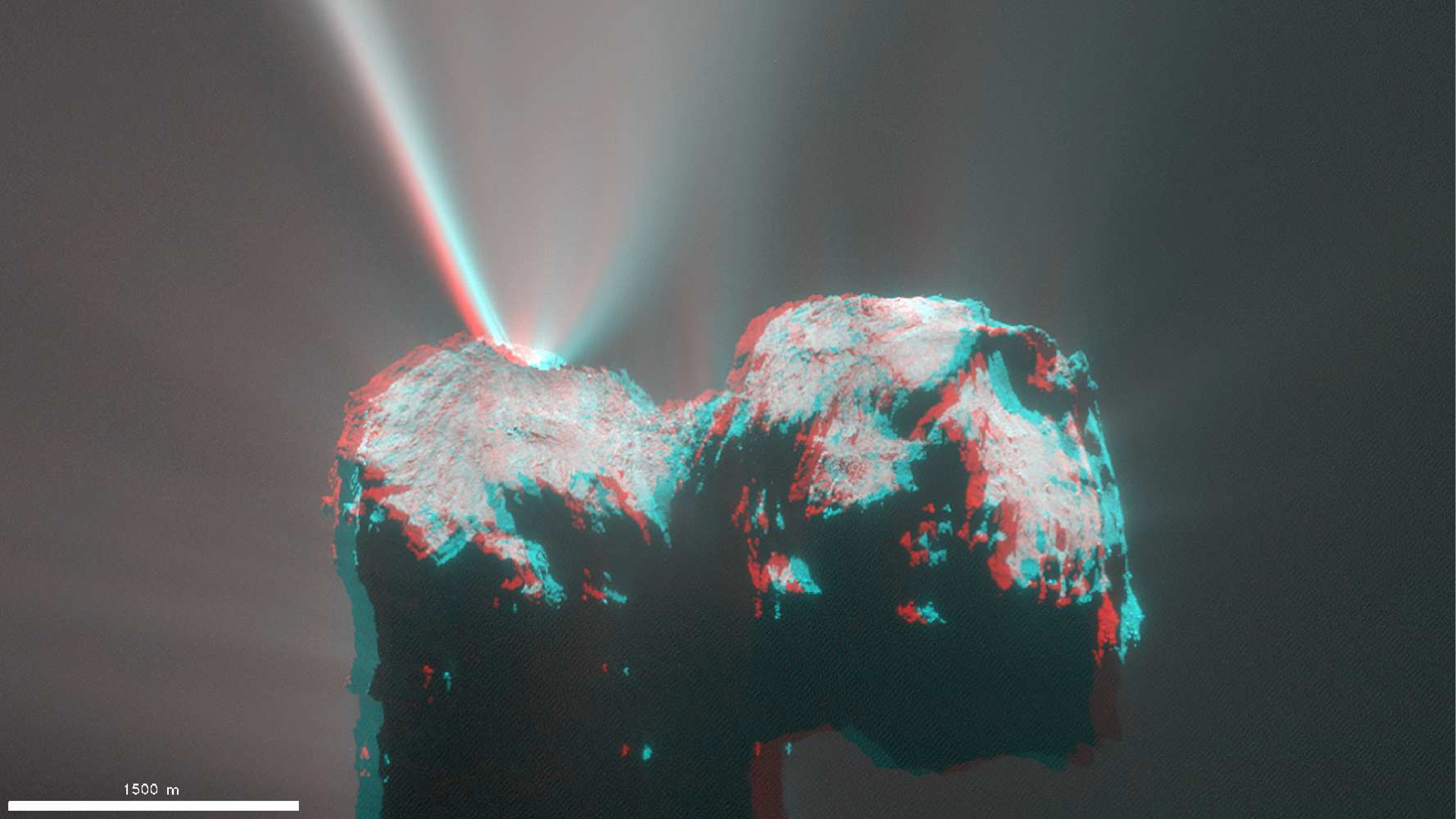}
   \caption{Anaglyphs of the nucleus of comet 67P/Churyumov-Gerasimenko after the powerful outburst of August 2015.
Note the impressive 3D rendering of the jets and remark the narrow, collimated jet next to the strong, more diffuse one on the left side of the fan.
A detailed analysis of the jet labeled ``14'' is presented in \cite{Vincent2016} }
   \label{anag_jet}
\end{figure*}

\begin{figure*}[htpb!]
   \centering
   \includegraphics[width=0.8\textwidth]{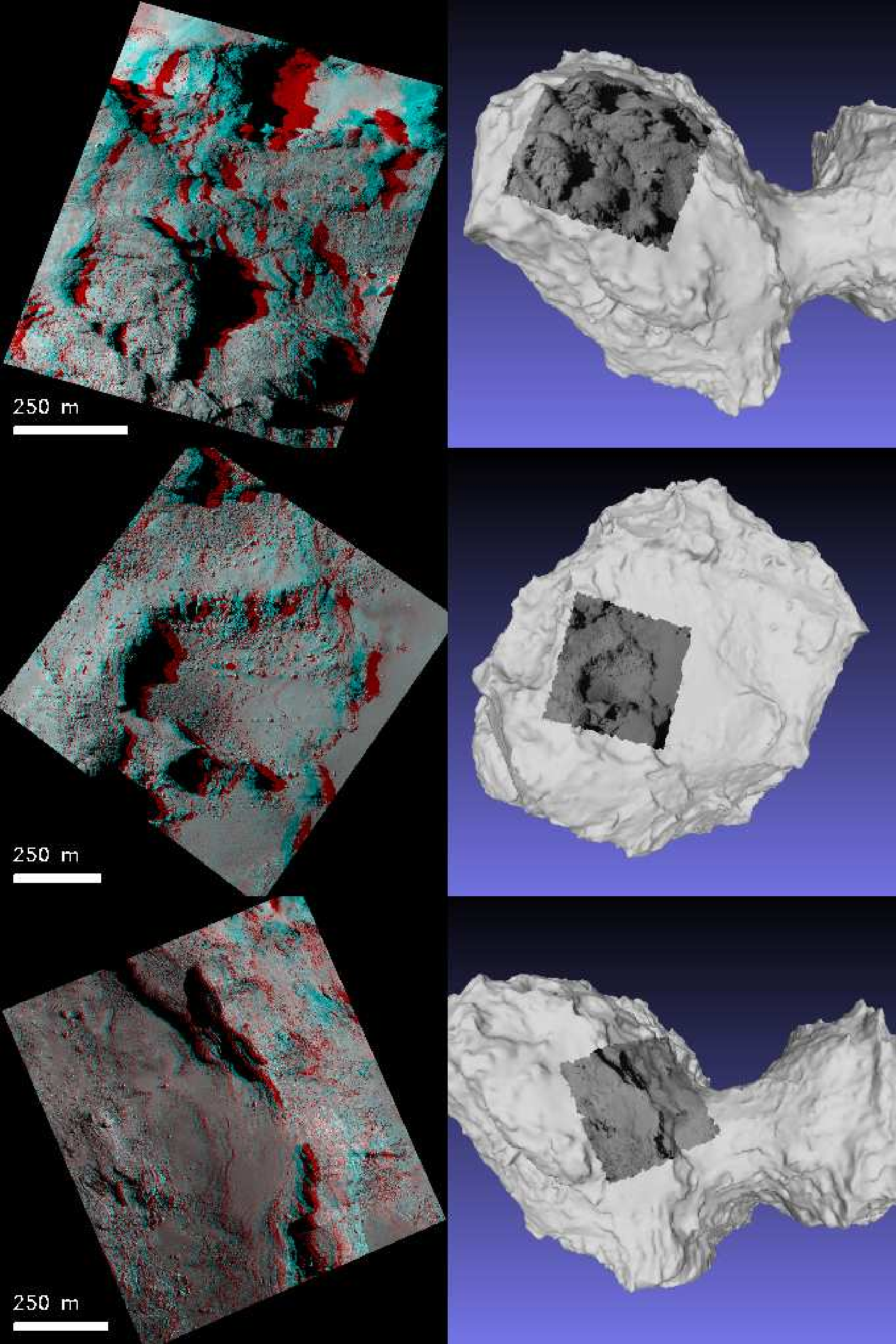}
   \caption{These three anaglyphs are located on the big lobe of the nucleus of comet 67P/Churyumov-Gerasimenko.
	   Top panel: this anaglyph extends over three regions, Anubis, Atum and Seth; it shows a superposition of eroded terraces in Anubis and Atume (lower part) and a small part of the Seth region characterized by a fine deposit (upper part). 
	   Middle panel: this anaglyph covers Imhotep and Khepry; it prominently shows a large, partially collapsed, basin on the border of Imhotep (central part). 
     Bottom panel: this anaglyph extends over four regions, Anubis, Atum, Hapi ans Seth; it shows the relatively flat southern regions of Anubis and Atum (left side) and the sharp discontinuity with the Hapi and Seth regions (right side)}.
   \label{mosaic_a}%
\end{figure*}

\begin{figure*}[htpb!]
   \centering
   \includegraphics[width=0.8\textwidth]{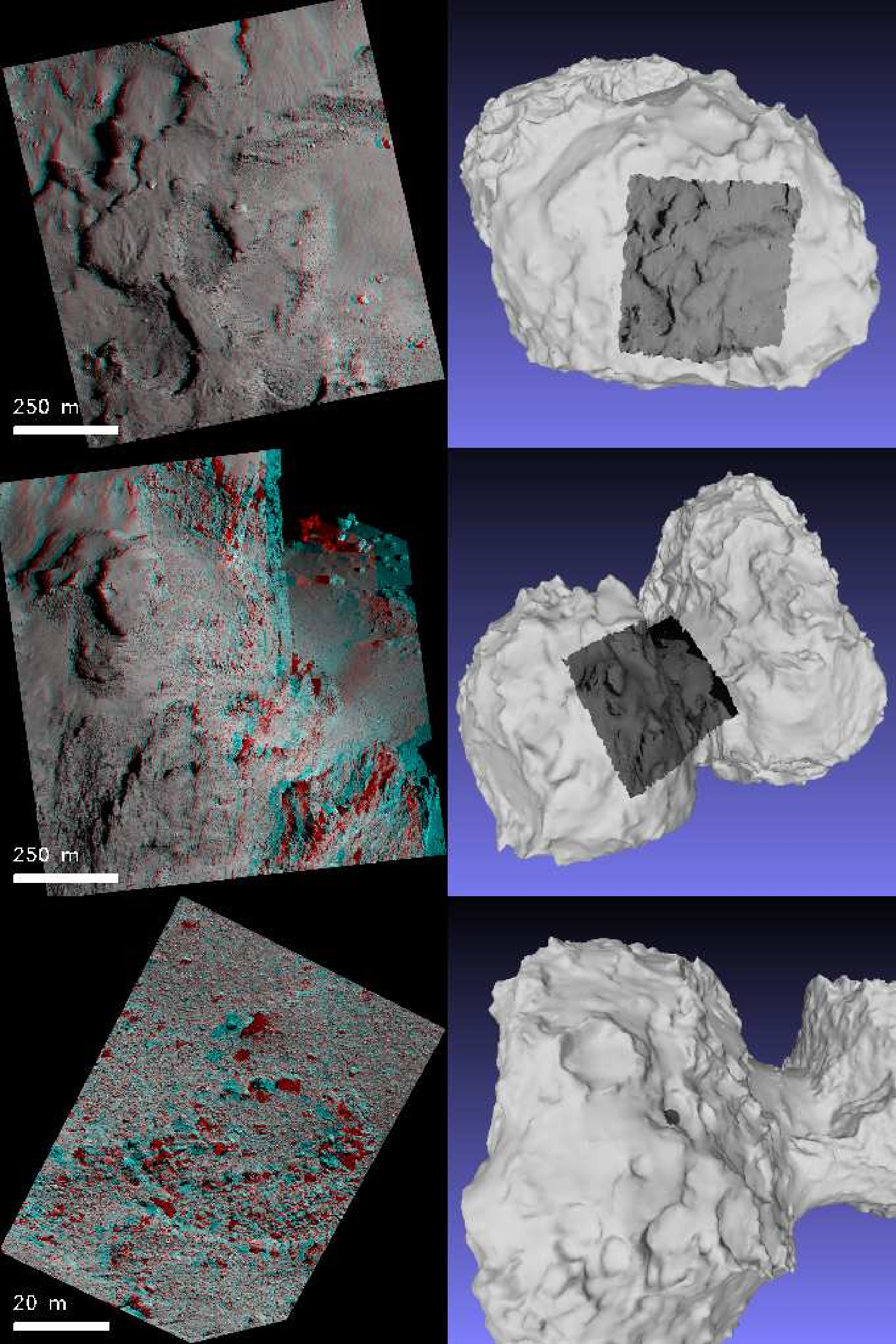}
   \caption{Top panel: this anaglyph is located on the small lobe of the nucleus of comet 67P/Churyumov-Gerasimenko and covers four regions, Hatmehit, Maat, Maftet and Nut; it shows the large Hatmehit depression with numerous boulder fields (lower right part) overlooking the Maat, Maftet and Nut regions extensively covered by a thick deposit.
     Middle panel: this anaglyph offers a dramatic perspective view over both the small lobe (Bastet and Maat) and the neck (Hapi).  
     Bottom panel: this anaglyph is located on the big lobe and covers a very small region (91 m$\times$65 m) of Seth; it shows a deeply eroded alcove that generated numerous meter size boulders over a field of cm-dm pebbles.}
   \label{mosaic_b}%
\end{figure*}

The anaglyphs of comet 67P/Churyumov–Gerasimenko implement the standard red/cyan system: the left image is coded in red levels and the right image is code in cyan (green+blue)) levels. 
Fusion of the two images is performed by the brain and the scene is seen in relief in gray levels thanks to the additive synthesis of colors. 
Careful attention must be given to the combination of the red/cyan glasses and the display screen to avoid crosstalk, that is leakages between the two color channels. 
This adverse effect limits the ability of the brain to successfully fuse the images perceived by each eye and thus reduces the quality of the 3D rendering. 
A simple, basic test can be performed by looking at the anaglyph with only one eye (blocking the other) and checking that a single image is seen (and not two slightly displaced images). 
For more comprehensive instructions, we direct the reader to the article by \citet{Woods2010} who quantified the crosstalk of a very large number of glasses and displays and to the website of co-author David Romeuf\footnote{\url{http://www.david-romeuf.fr/3D/Anaglyphes/BonCoupleEL/GoodCoupleMonitorGlassesAnaglyph.html}}.
The anaglyphs must be viewed in subdue light to fully grasp the weakest contrasts and the finest details.
White bands that may appear on both sides of the displayed anaglyph when it does not fill the screen must be avoided. 
Finally, most anaglyphs are of sufficiently quality for deep zooming thus offering dramatically detailed 3D views of the surface of the nucleus of comet 67P/Churyumov–Gerasimenko.
A few representative examples of anaglyphs are displayed in Fig.~\ref{anag_jet}, Fig.~\ref{mosaic_a}, and Fig.~\ref{mosaic_b}.
Regional information mentioned in the captions may be found in Fig.~\ref{6_views}.


\section{The parameters of the anaglyphs}

\begin{figure}[htpb!]
   \centering
   \includegraphics[width=0.5\textwidth]{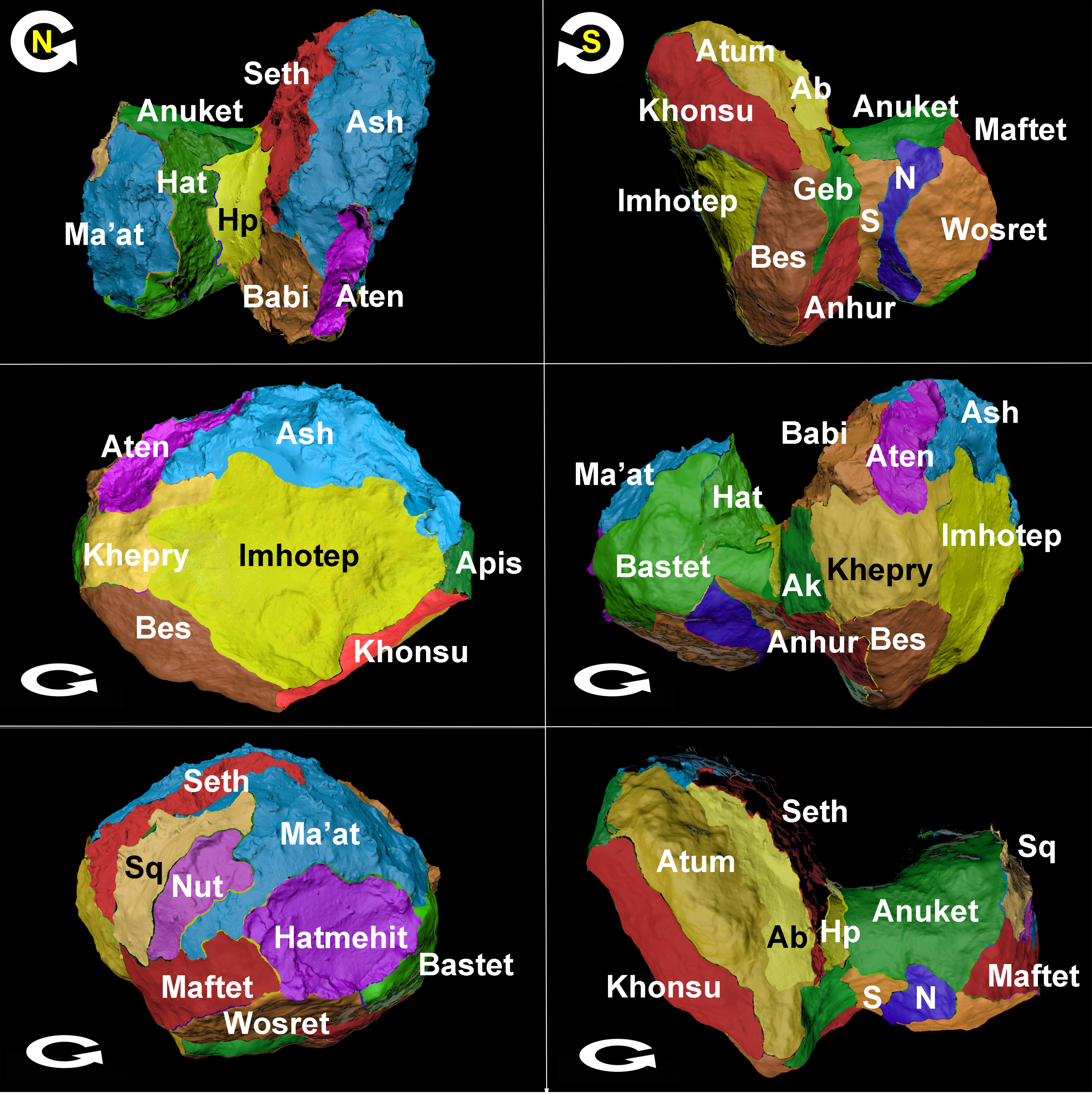}
   \caption{Six different views of the nucleus of 67P showing the 26 defined regions. 
The following names have been abbreviated for legibility: Hapi (Hp), Hathor (Hat) Sobek (S), Neith (N), Aker (Ak), and Serqet (Sq). Circular arrows show the direction of the comet’s rotation (reproduced from Fig. 2 of \citet{ElMaarry2017}).}
   \label{6_views}%
\end{figure}

Each anaglyph is documented by a set of 17 parameters which provide the contextual information. 
With the exception of the anaglyphs which show the whole nucleus, the spatial extent of the bulk of them is highly variable depending upon the distance to the comet. 
Particular attention was given to their localization on the nucleus. 
In addition to purely technical information (e.g. coordinates), different levels of localization are presented:

\begin{itemize} 
\item A global localization in terms of the main components of the nucleus, ``Big Lobe'' (BL), ``Small Lobe'' SL, and ``Neck'' (NK). 
Combinations are of course possible such as BL+NK.
\item A regional localization based on 26 morphological regions defined on the nucleus (\citet{ElMaarry2017}) and illustrated in Fig. ~\ref{6_views}.
\item A view of one of the image of the anaglyph projected on a 3D model of the nucleus as illustrated in Fig.~\ref{mosaic_a} and Fig.~\ref{mosaic_b}.
\end{itemize}

For the projection, a global 3D model of the nucleus resulting from a stereo-photogrammetric analysis of more than 1500 NAC images of the nucleus \citep{Preusker2017} was used. 
This model has 4 million facets and a spatial resolution of 3.4~m.

The name of an anaglyph concatenates the prefix “anag”, the name of the two images used for its creation, the value expressed by “P1sVAL” (VAL being the inverse of the parallax), and a code for the thresholding ``li'' for linear,``xp1s'' for square root). 
The parameters for each anaglyph listed below are part of the database and can therefore be queried in order to select a specific subset of anaglyphs.

\begin{itemize} 
\item Name of the first image (left/red image)
\item Name of the second image (right/cyan image)
\item Filter of the first image
\item Filter of the second image
\item Date and time of the earliest image used for the anaglyph
\item Global localization of the anaglyph: BL, SL, NK or combination
\item Regional localization of the anaglyph
\item Longitude of the center of the anaglyph (deg)
\item Latitude of the center of the anaglyph (deg)
\item Minimum distance to the closest pixel of the anaglyph (km)
\item Maximum distance to the farthest pixel of the anaglyph (km)
\item Mean distance obtained by averaging the distances of all pixels (km)
\item Mean spatial scale of the anaglyph (cm/pixel)
\item Size of the anaglyph as mean height and width of the anaglyph (m)
\item Phase (Sun-nucleus-Rosetta) angle (deg)
\item Parallax of the anaglyph
\item Feature of interest among the following seven items: Jets, Pits, Rings, Pancake, Agilka (the initial landing site of Philae), Abydos (the final landing site of Philae), and Philae.
\end{itemize}

\section{The on-line catalog}

\begin{figure*}[htbp!]
   \centering
   \includegraphics[width=0.9\textwidth]{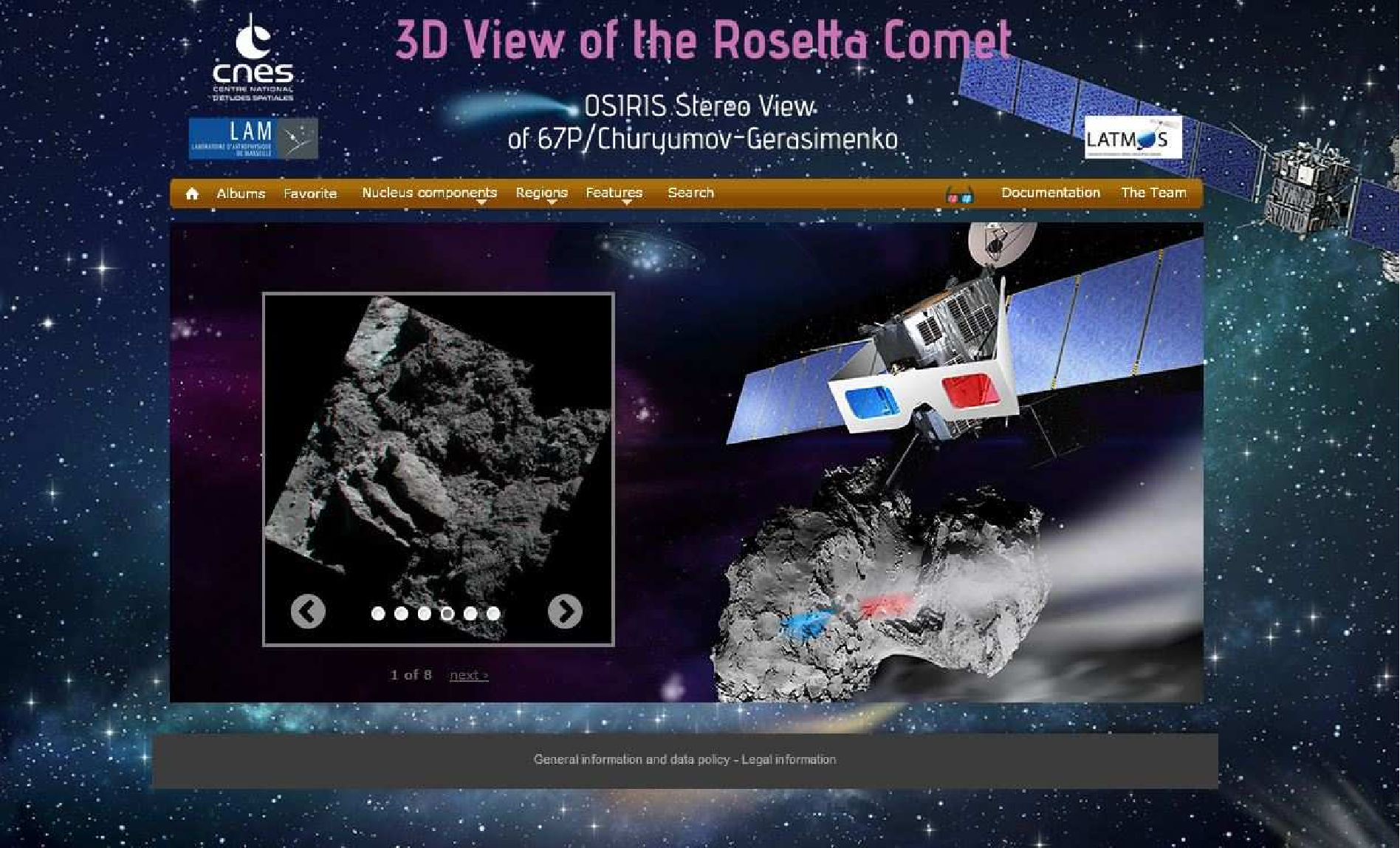}
   \caption{The home page of the dedicated website for anaglyphs access and visualization.} 
   \label{home_page}%
\end{figure*}

Figure~\ref{home_page} displays the home page of the dedicated website for anaglyphs access and visualization\footnote{\url{http://rosetta-3dcomet.cnes.fr}}, hosted by CNES (Centre National d'Etudes Spatiales/French Space Agency).
We describe below its main features.
The ``Favorite'' tab gives access to a subset of 55 anaglyphs selected for their interest and usually spectacular views of the nucleus.
The ``Nucleus components'' tab allows selecting anaglyphs covering the whole nucleus and its main components: BL, SL, and NK.
The ``Regions'' tab offers the same function but for the 26 regions of the nucleus.
The ``Features'' tab highlights four types of characteristic features of the nucleus (Jets, Pits, Rings, and Pancake), the original (Agilkia) and final (Abydos) landing site of Philae and views of Philae.
The ``Search'' tab allows searching specific anaglyphs according to:
\begin{itemize} 
\item their localization by either ``Nucleus components'', ``Regions'', or Longitude and Latitude,
\item the filters of the two images forming the anaglyphs,
\item the other parameters listed in the above section.
\end{itemize} 

\section{Conclusions}

Our project fits well with recent efforts that recognize the value of anaglyphs as a tool for the visualization and the characterization of solar system bodies. 
We are convinced that the systematic production of anaglyphs of comet 67P/Churyumov–Gerasimenko and their documented presentation on a dedicated website will contribute to future detailed analysis of its nucleus, will renew the interest of the general public, and will provide a valuable contribution to the legacy of the Rosetta mission.
Our catalog includes 1823 anaglyphs at all scales from the coma down to a few centimeters on the nucleus surface, thus probing its topography at scales unreachable by other stereo reconstruction technics.
In view of the success and interest of anaglyphs for solar system bodies, it is suggested that future planetary missions that are not equipped with stereo cameras carefully plan optimized observational sequences for proper anaglyph construction.

\begin{acknowledgements}
We are grateful to H. Gilardy for his assistance during the preparation of the manuscript.
The construction of the stereo anaglyphs and the realization of the catalog and associated website were funded by the Centre National d'Etudes Spatiales.
The International ROSETTA Mission was a cooperative project between ESA, several European national space agencies, and NASA.
\end{acknowledgements}

~\newpage

\bibliographystyle{aa}
\bibliography{Lamy_AA_Site-Anag_2018-09-14}

\end{document}